\documentclass[aps,prx,twocolumn,final,letterpaper]{revtex4}

\usepackage{appendix}
\usepackage{graphicx}   
\usepackage{import}                         
\usepackage{epstopdf}
\usepackage{amsmath} 
\usepackage{bm}
\usepackage{amssymb}
\usepackage{quotes}
\usepackage{transparent}
\usepackage{dcolumn}
\usepackage{multirow}
\usepackage{cancel} 
\usepackage{mdframed}
\usepackage{color}
\usepackage{bm}
\usepackage{dsfont}
\usepackage{slashed}
\usepackage{enumitem}
\begin{document}
\rmfamily

\title{Creating large Fock states and massively squeezed states in optics \\ using systems with nonlinear bound states in the continuum}

\author{Nicholas Rivera$^{1,2,\dagger}$, Jamison Sloan$^{3,\dagger}$, Yannick Salamin$^{3}$, John D. Joannopoulos$^{2}$, and Marin Solja\v{c}i\'{c}$^{2,3}$}

\affiliation{$^{1}$Department of Physics, Harvard University, Cambridge, MA 02138, USA.  \\
$^{2}$Department of Physics, Massachusetts Institute of Technology, Cambridge, MA 02139, USA.  \\
$^{3}$Research Laboratory of Electronics, Massachusetts Institute of Technology, Cambridge, MA 02139, USA.}

\begin{abstract} The quantization of the electromagnetic field leads directly to the existence of quantum mechanical states, called Fock states, with an exact integer number of photons. Despite these fundamental states being long-understood, and despite their many potential applications, generating them is largely an open problem. For example, at optical frequencies, it is challenging to deterministically generate Fock states of order two and beyond. Here, we predict the existence of a new effect in nonlinear optics, which enables the deterministic generation of large Fock states at arbitrary frequencies. The effect, which we call an \emph{n-photon bound state in the continuum}, is one in which a photonic resonance (such as a cavity mode) becomes lossless when a precise number of photons $n$ is inside the resonance. Based on analytical theory and numerical simulations, we show that these bound states enable a remarkable phenomenon in which a coherent state of light, when injected into a system supporting this bound state, can spontaneously evolve into a Fock state of a controllable photon number. This effect is also directly applicable for creating (highly) squeezed states of light, whose photon number fluctuations are (far) below the value expected from classical physics (i.e., shot noise). We suggest several examples of systems to experimentally realize the effects predicted here in nonlinear nanophotonic systems, showing examples of generating both optical Fock states with large $n$ ($n > 10$), as well as more macroscopic photonic states with very large squeezing, with over 90\% less noise (10 dB) than the classical value associated with shot noise. 
\end{abstract}

\maketitle


The principle of wave-particle duality is at the core of the modern understanding of the electromagnetic (EM) field. Central to the particle side of the duality is the idea of quantization, which states that the EM field is composed of discrete packets of energy (photons). In the language of quantum electrodynamics, this quantization is expressed by the Fock states $|n\rangle$ ($n = 0, 1, 2, \dots$), which are eigenstates of the photon number operator $a^{\dagger}a$, where $a$ is the annihilation operator of a quantized mode of the electromagnetic field. Being photon number eigenstates, they have an exactly defined integer photon number, with zero uncertainty. They are also eigenstates of the electromagnetic energy operator (the Hamiltonian), $H = \hbar\omega a^{\dagger}a$, where $\hbar$ is the reduced Planck constant, and $\omega$ is the frequency of the EM mode in question. As the eigenstates of the Hamiltonian, they are also time-harmonic solutions to the Schrodinger equation for the electromagnetic field. In that sense, Fock states are the most elementary quantum states of light, and as such, have played a foundational role in our understanding of the quantum theory of light.

Accordingly, Fock states have long been identified by the quantum community as an important state to access. For example, these highly nonclassical states have long been considered for precision measurements (in the field of quantum metrology) because they have no uncertainty in their photon number (or equivalently, their intensity). These states are thus not subject to the so-called shot noise associated with the Poissonian fluctuations of photon number in classical light pulses (corresponding to quantum-mechanical coherent states) \cite{teich1989squeezed,davidovich1996sub,thomas2011real}. One of the earliest proposed applications in precision measurement was to use a Fock state in an optical cavity as an extremely sensitive sensor of small vibrations \cite{braginskiui1975quantum}. Beyond applications in precision measurements, they are also considered valuable for the fields of quantum simulation and quantum information processing. For example, Fock states of microwave frequency fields in LC resonators have already been used for quantum chemistry tasks, such as calculating the energy spectra of molecules \cite{wang2020efficient}. Another high-profile application of large Fock states along these lines is in the implementation of quantum algorithms such as boson sampling \cite{aaronson2011computational,lund2014boson,huh2015boson,wang2017high,hamilton2017gaussian, brod2019photonic}, in which the passage of a Fock state through an array of waveguides and beamsplitters can be used to calculate matrix permanents. For this application, a “modest” Fock state of even 100 photons can enable computations of matrix permanents at least fifteen orders of magnitude larger than could be handled by even the largest supercomputers today \cite{zhong2020quantum}. Beyond this, Fock states are also important ``resource states,'' which enable the generation of other desirable quantum states. For example,  Fock states can be used to generate Schrodinger cat states \cite{ourjoumtsev2007generation},  displaced Fock states, and Gottesman-Kitaev-Preskill states,  which all have been identified as having important applications in quantum computation. 

For the reasons above and many more, the problem of generating Fock states has generated considerable attention. At microwave frequencies, it is possible to deterministically produce Fock states of modest sizes (roughly 15 photons), with a high rate of success ($> 90\%$). These Fock states are generated in microwave resonators through a combination of external driving of the cavity by microwave pulses and superconducting transmon qubits \cite{heeres2017implementing} (which provide a very strong nonlinearity). Fock states have also been generated in microwave cavities by strongly coupling them to transmon qubits that are repeatedly pumped to inject photons into the cavity at deterministic times \cite{hofheinz2008generation}. Older foundational work in the field of cavity quantum electrodynamics made use of Rydberg atoms strongly coupled to microwave cavities in order to generate low-order Fock states using principles such as the one above, as well as quantum feedback protocols \cite{ rempe1990observation, varcoe2000preparing, sayrin2011real}. Such Rydberg atom-cavity interactions form the basis for new theoretical proposals to extend microwave Fock states to higher photon numbers \cite{uria2020deterministic, canela2020bright}. 

\begin{figure}[t]
    \centering
    \includegraphics[width=0.475\textwidth]{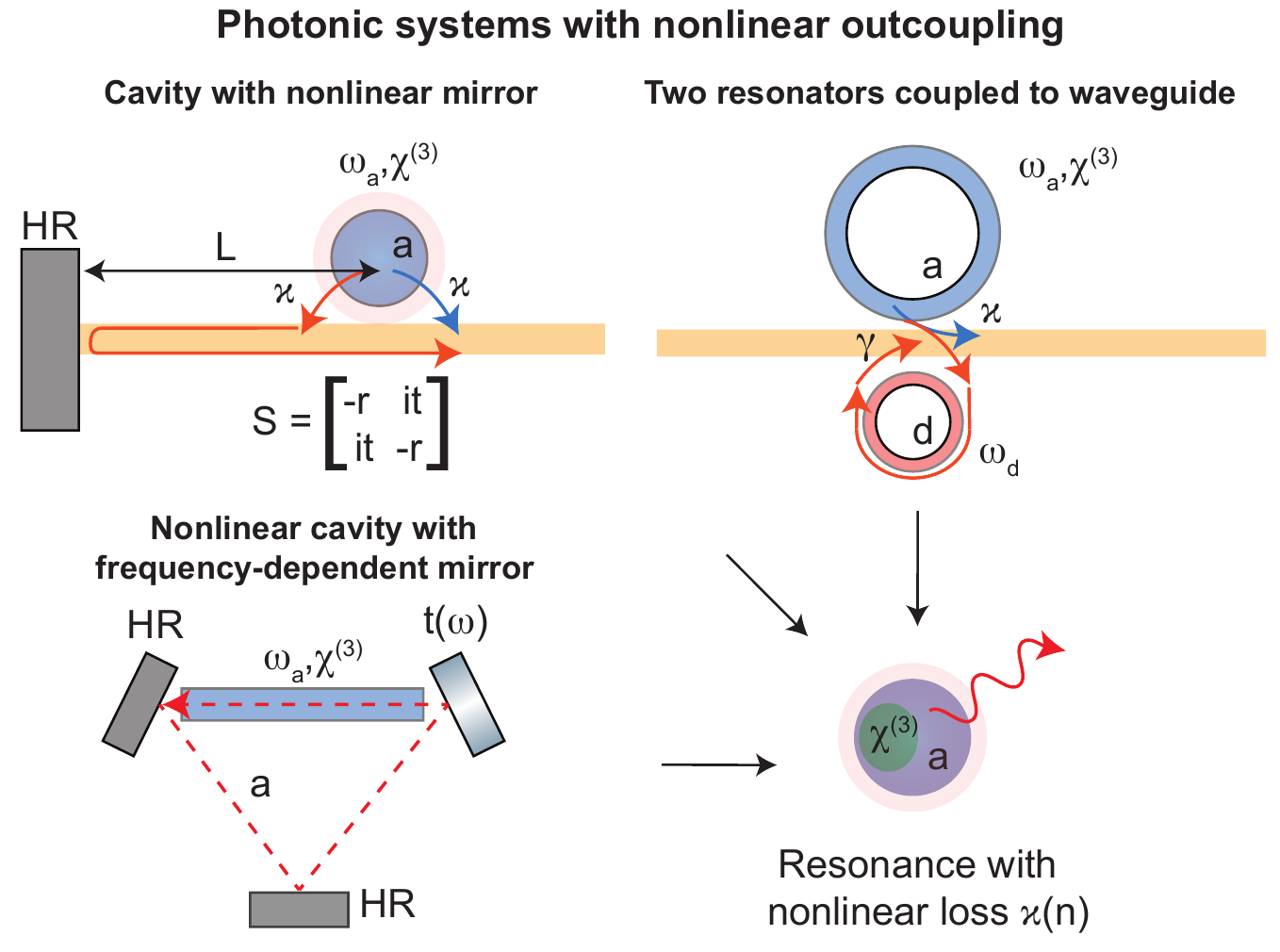}
    \caption{\textbf{Examples of photonic resonators with nonlinear radiation loss.} (Top left) A high-Q resonator (blue) of frequency $\omega_a$ with third-order nonlinear susceptibility $\chi^{(3)}$, which is coupled to a waveguide (yellow) that is terminated by a perfect reflector (gray, denoted HR) at a length $L$ away from the resonator. The resonator can leak out directly to the right at rate $\kappa$ (blue path) or reflect from the left mirror (orange path) before transmitting to the right (with direct reflection and transmission coefficients $r,t$). The relative phase of the interference between blue and orange leakage paths depends on $\omega_a$, leading to a frequency-dependent $Q$ factor which can be infinite when the two paths destructively interfere. (Top right) Another example of the same physics is realized by two resonators (denoted $a$ and $d$) coupled to a common waveguide, one with rate $\kappa$ (blue) and the other with rate $\gamma \gg \kappa$ (orange). Light from the high-Q (blue) resonator can leak out directly, or passing through the low-Q resonator first. The interference depends on the two resonance frequencies, denoted $a, d$ and can in some cases completely cancel. (Bottom left) A cavity formed by two highly reflective (HR) mirrors and one mirror with frequency-dependent transmission $t(\omega)$ can also have a nonlinear loss when a nonlinear medium is integrated inside of it. (Bottom right) All of these three examples can be thought of as a single resonator with a loss rate for photons $\kappa(n)$ that depends on the number of photons, $n$, inside the resonance.}
\end{figure}

In optics, the situation is drastically different, where the techniques above cannot be applied. In fact, it is currently a significant challenge to deterministically produce Fock states in optics when $n \geq 2$. This lies in stark contrast to single photon Fock states, where the capabilities are much more mature \cite{lounis2005single}. In general, proposals for generating the Fock state $|n\rangle$ ($n \geq 2$) are either non-deterministic (meaning the state is generated with random order, or at random times \cite{yamamoto1987generation,waks2004direct, cooper2013experimental, ben2021shaping}), or are otherwise rather complex and resource intensive \cite{bonifacio1971quantum, bonifacio1971quantum2, gonzalez2017efficient,kilin1995fock,leonski1997fock, yanagimoto2019adiabatic}, requiring highly correlated (superradiant) states of many-atoms, exotic high-order photonic nonlinearities, and/or extremely complex pump protocols $-$ all of which have thus far not been implemented.

Here, we predict a remarkable new effect in nonlinear optics that enables deterministic creation of large Fock states. In particular, Fock states result from simply: taking a \emph{specially designed photonic resonator with an intensity-dependent refractive index}, injecting laser light into that resonator (which loads a coherent state into it), and letting the photons decay from the resonator through radiation loss (leakage of light from the resonator). Radiation loss in this special resonator will $-$ in the ideal case of the effect we predict $-$ convert the coherent state into a Fock state with perfect efficiency. This is in stark contrast to what happens normally in resonators with loss, where classical (coherent) states stay classical upon their decay, and nonclassical states become classical. Here a classical state, after decaying, becomes extremely nonclassical (a large Fock state). This surprising result is accounted for by the fact that the specially designed resonator we introduce supports a \emph{bound state of many photons}: namely the resonator is \emph{dissipationless} when there are $n$ and exactly $n$ photons in the resonator and it is dissipative otherwise.  

We show that new many-photon bound state, which we call an \emph{$n$-photon bound state in the continuum}, also facilitates generating very strongly squeezed states of photons. Squeezed states, in particular photon number-squeezed states, are quantum mechanical states of the radiation field with photon number variance $(\Delta n)^2$ less than the mean $\langle n \rangle$: $(\Delta n)^2 < \langle n \rangle$. In other words, their fluctuations are below the value associated with the shot noise typical of classical light (in quantum mechanical coherent states). Because shot noise is associated with Poissonian fluctuations of the photon number, such photon-number squeezed states with variance below the mean are often called sub-Poissonian. Fock states, with $\Delta n = 0$ are a special case of number-squeezed states, representing the ultimate degree of squeezing. Number-squeezed states are very useful and desirable in their own right for precision measurement and have been thoroughly explored in nonlinear optics \cite{schmitt1998photon, shirasaki1990squeezing,bergman1991squeezing,kitching1994amplitude, friberg1996observation, andersen201630}. There is a large push to realize strong squeezing in nanophotonic systems, which would open up many applications in quantum sensing and precision measurement. That said, large degrees of squeezing (e.g., 10 dB) have proven elusive, although impressive strides have been made recently based on the use of second-order nonlinearities \cite{nehra2022few}.  The effects we introduce enable very large degrees of squeezing in nanophotonic systems, and we show examples of how photon-number fluctuations can realistically be reduced by over 90\% (10 dB) of the classical shot noise level.

Before moving to the results, we note that these remarkable predictions are the direct consequence of a new concept that we put forth in this paper: photonic structures with nonlinear radiation loss (Fig. 1). In other words, photonic structures where the \emph{lifetime} of the photons depends on how many photons are inside the structure. We expect this concept to have many applications even for classical optical devices such as lasers, modulators, sensors, and switches. Here, we focus on the remarkable \emph{quantum} effects that light experiences when leaking out of these structures, developing the first-principles quantum optics theory of them. It must be emphasized it is precisely due to this nonlinear radiation loss that we are able to predict, for the first time, the possibility of directly producing Fock states by direct laser excitation of simple and existing nonlinear optical structures.

\section{Radiation loss in nonlinear photonic resonances} 

We begin by describing the new effects and the intuition behind them. The physics we are interested in is that of radiation loss in photonic resonators made with materials that have an intensity-dependent refractive index (Kerr nonlinearity). We consider a special type of photonic resonator, of which Fig. 1 shows three instances. What all three structures have in common is that they can have very high $Q$-resonances due to destructive interference between two or more ``paths'' for light in the resonator (labeled $a$) to escape to the continuum. 

Such systems, where high-$Q$ arises from destructive interference between radiation loss pathways, have been the subject of many recent works in the photonics community, under names like \emph{bound states in the continuum} (BICs) \cite{hsu2016bound}, \emph{quasi-bound states in the continuum}  \cite{rybin2017supercavity,koshelev2018asymmetric}, and \emph{Fano resonances} \cite{limonov2017fano}. Characteristic to these high-$Q$ resonances is that their $Q$-factor sensitively depends on the geometrical and material parameters of the resonator (e.g., feature size, index of refraction, photon wavevector),and the $Q$ achieves a very large maximum for some value of these parameters. This maximum occurs for the geometrical parameters which lead to opposite phases for the two leakage paths (e.g., blue and orange paths in the first and second systems of Fig. 1). When this happens, the $Q$ is limited only by what we'll call ``external'' losses (e.g., absorption, scattering, added outcouplers, etc.). As a point of terminology, we will generally use the term BIC to refer to ``cancellation-induced'' high-$Q$ resonances, in keeping with current usage of the phrase \cite{rybin2017supercavity,yu2021ultra} \footnote{Despite many of these structures being theoretically unable to achieve literally infinite quality factor, due to their finite extent \cite{hsu2016bound})}.

Now, we consider the quantum mechanical dynamics of the intraresonator photon state, due to the leakage of light out of a resonator with a BIC and Kerr nonlinearity. At first glance, it seems that the dynamics of radiation loss from these nonlinear high-$Q$ resonances would just be governed by the textbook theory of dissipation in nonlinear high-$Q$ resonators \cite{drummond1980quantum,gardiner2004quantum,walls2007quantum}. The conventional theory has been applied extensively for over forty years, predicting a variety of effects which have been observed, such as optical bistability (in the classical domain \cite{haus1984waves}), dissipative phase transitions \cite{fink2018signatures}, and modest amplitude squeezing (antibunching of light) in the cavity mode \cite{boulier2014polariton,delteil2019towards,munoz2019emergence}. This natural assumption, that it is only the value of $Q$ that matters, is surprisingly not correct.

Consider what happens when we add Kerr nonlinearity to the resonance $a$ (for example, when the medium of the cavity in Fig. 1 (left) has third-order optical nonlinearity). In that case, the refractive index depends on intensity, or equivalently, the number of photons, $n$, in the cavity. Then, as the intracavity photon number changes, so does the resonance frequency $\omega_a$ \cite{joannopoulos2011photonic}, and so do the relative phases of leakage paths in Fig. 1a. In such a case, the $Q$-factor depends on the number of photons in the cavity: $Q = Q(n) = Q(\omega_a(n))$: it is the composition of the dependence of $Q = Q(\omega_a)$ on the resonator frequency (amplitude-phase coupling) and the dependence of the resonator frequency $\omega_a = \omega_a(n)$ on the photon number (because of the Kerr nonlinearity). This system therefore has \emph{nonlinear loss}: a decay rate $\kappa(n)$ that depends on the number of photons in $a$. 

We will show that Kerr nonlinear BICs realize a very unique form of nonlinear loss for photons compared to well-known forms of nonlinear loss, like saturable or multiphoton absorption. This is illustrated in Fig. 2, where we plot the nonlinear loss rate as a function of photon number (parameters will be explained in the section on proposed experimental realizations). As anticipated from the arguments above, there is a maximum in $Q$ (minimum in $\kappa$) as a function of photon number. For the case of an ideal BIC (infinite lifetime), there is a special photon number $n_0$ for which the loss is exactly zero $\kappa(n_0) = 0$. We refer to the structure as having an \emph{$n_0$-photon BIC}, because when it has $n_0$ photons, the resonance lifetime is infinite.

\begin{figure}[t]
    \centering
    \includegraphics[width=0.475\textwidth]{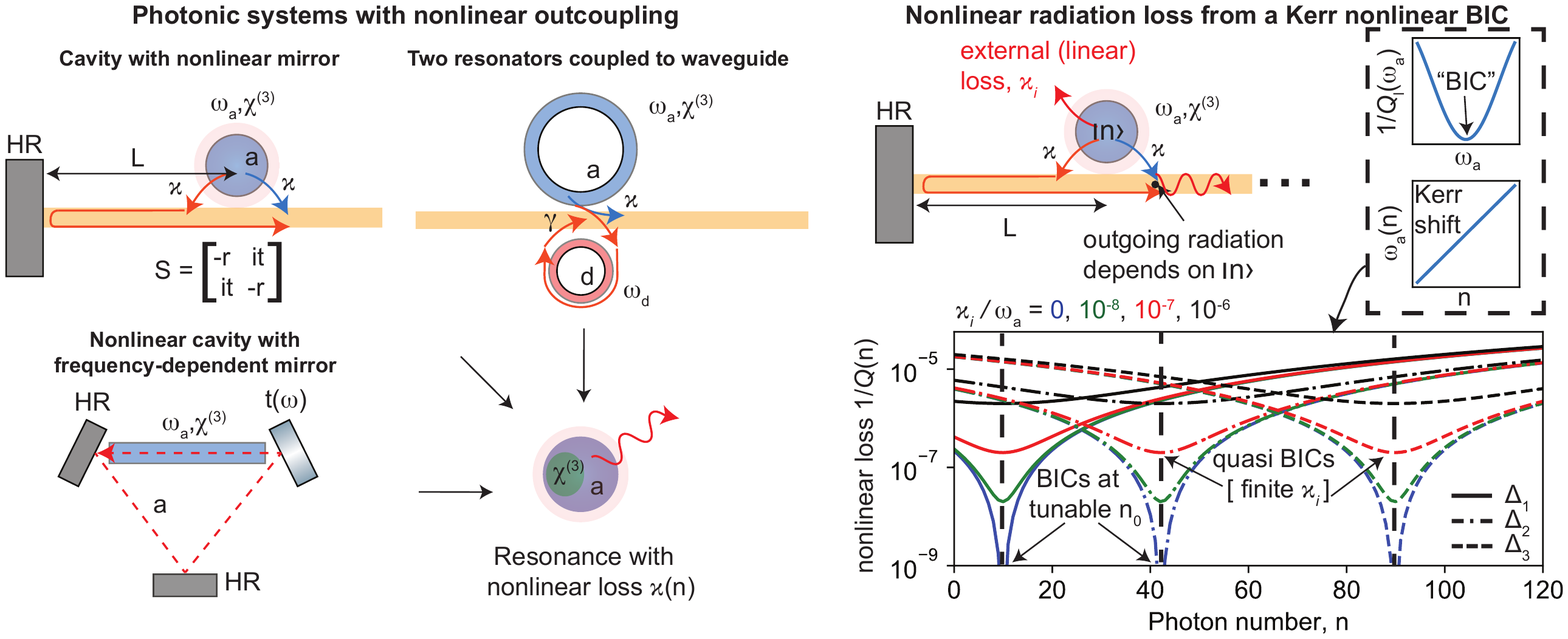}
    \caption{\textbf{n-photon bound states in the continuum.} Calculated nonlinear loss for the structure in the top left of Fig. 1. Due to destructive interference between two paths for radiation (blue and orange), it is possible for the coupling of the resonator to the waveguide to vanish for a certain resonator frequency $\omega_a$, leading to a minimum in the loss (top inset, showing the inverse quality factor $Q^{-1}(\omega_a)$). Due to the nonlinear refractive index (Kerr nonlinearity), the resonator frequency depends on the intracavity photon number, $\omega_a = \omega_a(n)$ (bottom inset), leading to a photon-number dependent loss for the cavity. In the case of a so-called \emph{bound state in the continuum} (BIC), where the lifetime of the resonance becomes infinite, the loss becomes zero for some photon number $n_0$ $-$ what we call a structure with an $n_0$-photon BIC. When there is finite external linear loss, denoted by $\kappa_i$, the nonlinear loss reaches a sharp minimum, rather than a zero, for that same photon number. Changing the detuning of the resonator from the frequency of minimum loss changes the zero-loss-photon-number $-$ here this is shown for three different detunings of roughly $\Delta_{1,2,3} = (0.0094,0.043,0.093)\gamma$, where $\gamma = 2L/v_g$, with $v_g$ the group velocity of the waveguide. The parameters used in this figure are: $\omega_a = 1.47$\text{eV}, $\gamma = 10^{-2}\omega_a$, $\kappa = 10^{-3}\omega_a$, $\beta = 5\times 10^{-6}\omega_{a}$. Detailed experimental justification of parameters is provided in the discussion of Fig. 5 and SI Section IV.}
    \label{fig:fock1}
\end{figure}
It is clear that such a system might facilitate creation of $n_0$-photon Fock states. Suppose we populate the system with an average number of photons $\bar{n}$ greater than $n_0$ (for example, by populating the resonance with a coherent state with a mean of $\bar{n}$ photons and a variance of $\bar{n}$ photons). Then the system will decay until it has $n_0$ photons exactly (with zero variance): the system will stay stuck in the state $|n_0\rangle$ because the loss rate is zero in that state, and the photons have nowhere to go at that point; the variance (fluctuations) have disappeared entirely. By continuity, even when the BIC is imperfect (as it always is) due to some finite external loss rate, denoted $2\kappa_i$ (see Fig. 2), one expects the variance in photon number to rapidly drop (well below the mean), leading to the photon-number-squeezed light described earlier. \emph{Thus, even though the BIC is never perfect, we will find that impressively good approximations can still emerge in realistic settings.} That said, if the background loss is too high, these quantum optical effects will be washed out, as the nonlinear loss looks linear (independent of photon number): see the black curves of Fig. 2, where one would expect the usual physics of dissipation in which nonclassical states become classical.

\subsection{Theory of nonlinear radiation loss}
In what follows, we show concrete examples and applications of this concept, showing that it remains interesting in realistic cases. The results arise from a general theory that we develop, of dissipation in nonlinear photonic structures with frequency-dependent (``non-Markovian'') outcouplings. We emphasize that this new analytical theory we develop, along with its remarkable consequences, have also been verified (SI) by standard numerical  simulation tools for open-quantum-systems. First, we state the key results of this theory (see SI for detailed derivations). The family of systems to which our theory immediately applies is a driven Kerr nonlinear resonance, coupled to some number ($N$) of reservoirs (also called continuua; they can be associated with radiation or absorption loss \footnote{While the origin of the continuum modes can be arbitrary, we will focus on the case in which the resonator is coupled to a continuum of far-field radiation modes.}). Coupling of the resonance to the reservoir leads to dissipation of the resonance. For this class of systems, the Hamiltonian takes the form:
\begin{align}
    &H/\hbar = \omega_a a^{\dagger}a + \beta\omega_a a^{\dagger 2}a^2 + \alpha(t) a + \alpha^*(t) a^{\dagger} \nonumber \\ &+ \sum\limits_{i=1}^{N}\int \frac{d\omega}{2\pi}~ \omega s_i^{\dagger}(\omega)s_i(\omega)  \nonumber \\ &+ \sum\limits_{i=1}^{N}\int\frac{d\omega}{2\pi}~i \left(K_{c,i}(\omega)s_i(\omega)a^{\dagger} - K_{c,i}^*(\omega)s^{\dagger}_i(\omega)a\right).
\end{align}
Here, $a$ is the annihilation operator of the resonance with frequency $\omega_a$. The Kerr nonlinearity of the resonance is well-known to manifest as the second term in the Hamiltonian \cite{drummond1980quantum,walls2007quantum}, leading to a ``photon-number-dependent resonance frequency.'' Given $n$ photons in the resonator, the energy to add another is $\hbar\omega_{n+1,n} \equiv \hbar\omega_a(1+2\beta n)$, where $\beta$ is a (dimensionless) nonlinear coefficient. The coherent-drive strength is $\alpha(t)$ and is not restricted to be monochromatic. The continuum modes are described as usual by a set of harmonic oscillators \cite{scully1999quantum}: for the $i$-th reservoir, the continuum modes are labeled by their frequency $\omega$ and annihilation operator $s_i(\omega)$. 

The resonance couples to the continuum through the last term in the Hamiltonian with coupling coefficient $K_{c,i}(\omega)$. This $K_{c,i}$ is the in-coupling function from the $i$-th reservoir: given a monochromatic input (at frequency $\omega$) from reservoir $i$ into the resonator, the classical amplitude of the resonance, denoted $a(\omega)$, is proportional to $K_{c,i}(\omega)$. It can be directly calculated from numerical electromagnetic simulations, for example, by launching an input wave at a resonance in a simulation and examining the frequency-domain amplitude of the resonance. This makes our theory \emph{ab initio} and broadly applicable. The in-coupling function can also be extracted from temporal coupled mode theory models \cite{haus1984waves,fan2002analysis, fan2003temporal}, which are well-known to accurately describe many widely-used photonic architectures. In the SI, we provide explicit forms of $K_c(\omega)$ for a few common photonic architectures that have BICs (the BIC is encoded in a zero/minimum of $K_c(\omega)$ as a function of frequency).

The Hamiltonian of Eq. (1) is exact, but not generally solveable. Now, we employ the sole approximation of our theory: that the bandwidth of the reservoirs, $\Delta\omega_i$ is much larger than the inverse ``response'' timescale of the reservoir $T^{-1}$. The physical quantity which sets $\Delta\omega_i$ and $T^{-1}$ varies from architecture to architecture, but it very frequently holds. This ``semi-Markovian'' approximation \cite{gardiner1987input, gardiner2004quantum} yields an equation of motion for the reduced density matrix of the cavity, denoted $\rho$. It is (see SI):
\begin{equation} 
    \label{eq:nonlinear_master}
    \dot{\rho} = -i[H_K + H_{\text{drive}},\rho] + \mathcal{D}[\rho]
\end{equation}
where $H_K + H_{\text{drive}} = \omega_a a^{\dagger}a + \beta\omega_a a^{\dagger 2}a^2 + \alpha(t)a + \alpha^*(t)a^{\dagger}$ is responsible for the conservative parts of the evolution of the resonance, and the dissipator $\mathcal{D}$ is defined through its matrix elements as ($m,n$ are Fock states):
\begin{align} 
    &\langle m| \mathcal{D}[\rho] |n\rangle  = -\left(mK_l(\omega_{m,m-1}) + nK^*_l(\omega_{n,n-1}) \right)\rho_{m,n} + \nonumber \\
    &\sqrt{(m+1)(n+1)}\left(K_l(\omega_{m+1,m})+K^*_l(\omega_{n+1,n}) \right)\rho_{m+1,n+1} .
\end{align}
The function $K_l(\omega)$ is the ``loss function'' or ``frequency-dependent loss'' of the cavity, and is directly connected to the incoupling function by a Kramers-Kronig relation, as:
\begin{equation}
    K_l(\omega) = \sum_i K_{l,i}(\omega), \text{with } K_{l,i}(\omega) \equiv i\int \frac{d\omega'}{2\pi }\frac{|K_{c,i}(\omega')|^2}{\omega-\omega'+i\eta}, 
\end{equation}
where $\eta$ is an infinitesimal. The function $K_l$ is related to the ``frequency-dependent $Q$ factor'' of Fig. 2 by $Q^{-1}(\omega) = 2\text{Re }K_l(\omega)/\omega$. The relation of Eq. (3) enforces the intimate connection between leakage and incoupling. Equations (2-4) summarize our quantum theory of nonlinear outcoupling: they tell us that the evolution of the quantum state of a resonance, including photon probabilities, field correlations ($g^{(1)}$), intensity correlations ($g^{(2)}$), and so on $-$ are strongly controlled by the photon number ($n$) dependence of quantities like: $K_l(\omega_{n+1,n}) = K_l(\omega_a(1+2\beta n))$, whose form can be controlled by engineering the outcoupling of light in a cavity. The theory here differs from the standard theory of leaky resonators, which typically makes an approximation called a ``white noise'' approximation. In this approximation, one neglects the frequency-dependence of the continuum coupling, thus neglecting the possibility that the resonator $Q$ depends on its frequency, as it does for the resonators of Fig. 1. \footnote{Our theory is mapped to the standard theory of damping of a resonator with the identification $K_c(\omega) = \sqrt{2\kappa}$, with $\kappa$ the amplitude decay rate (yielding $K_l = \kappa$ and reducing Eq. (2) to the standard master equation for a damped cavity as in \cite{scully1999quantum})}.

Let us now discuss what dictates the value of $n_0$ which stabilizes the Fock state. The $n$-photon BIC condition, in the language of our theory, is
\begin{equation}
\text{Re } K_l(\omega_{n_0,n_0-1}) = 0.
\end{equation}
Suppose the zero of the loss function $K_l(\omega)$ occurs at some frequency $\omega_{\text{0}}$ (the BIC frequency). Then, we may expand $K_l$ around $\omega_0$ as $\text{Re }K_l(\omega) \approx c_2(\omega-\omega_0)^2$ \footnote{There is no linear component because loss has to be non-negative.}. From Eq. (5), we have that 
\vspace{-0.05in}
\begin{equation}
    n_0 = \frac{\Delta_0}{2\beta} + 1,
\end{equation}
where $\Delta_0 \equiv \omega_0 - \omega_a$ is the detuning of the linear resonance from the BIC frequency. This simple equation shows that the order of the Fock state can be controlled by simply tuning the resonator frequency (see Fig. 2), and that there are discrete detunings that lead to Fock states. This equation also reveals that larger single-photon nonlinearities ($\beta$) and smaller detunings ($\Delta_0$) lead to smaller photon numbers, while smaller single-photon nonlinearities (characteristic of ``bulk material'' nonlinearities) lead in principle to Fock states at larger photon numbers. The larger Fock states are more fragile, but we will see that even so, intense and extremely squeezed light can be realized $-$ beyond what is typically achievable in resonators through normal nonlinear loss $-$ and thus this regime is still very interesting.

The dynamics of various quantum states undergoing the nonlinear radiation loss of Fig. 2 are illustrated in Fig. 3, in a ``phase space plot'' where the two variables being plotted are mean and variance.  Each line indicates a trajectory of some initial state: we show the dynamics of a set of initially ``Poissonian'' states with shot noise photon number fluctuations, as well as initial Fock states. Let us consider the dynamics of the Poissonian states with mean greater than $n_0 = 10$. For a true BIC (top panel), the trajectories move towards a Fock state of order 10 (blue circle in top panel). States with mean sufficiently below $n_0 = 10$ decay to vacuum, as expected. Intriguingly, initial states that have significant probability to be both at $n < 10$ and $n \geq 10$ lead to a ``mixed'' state that has some probability of being in vacuum and some of being a Fock state (see trajectories terminating for example at the purple point in Fig. 3).

\begin{figure}[t]
    \includegraphics[width=0.475\textwidth]{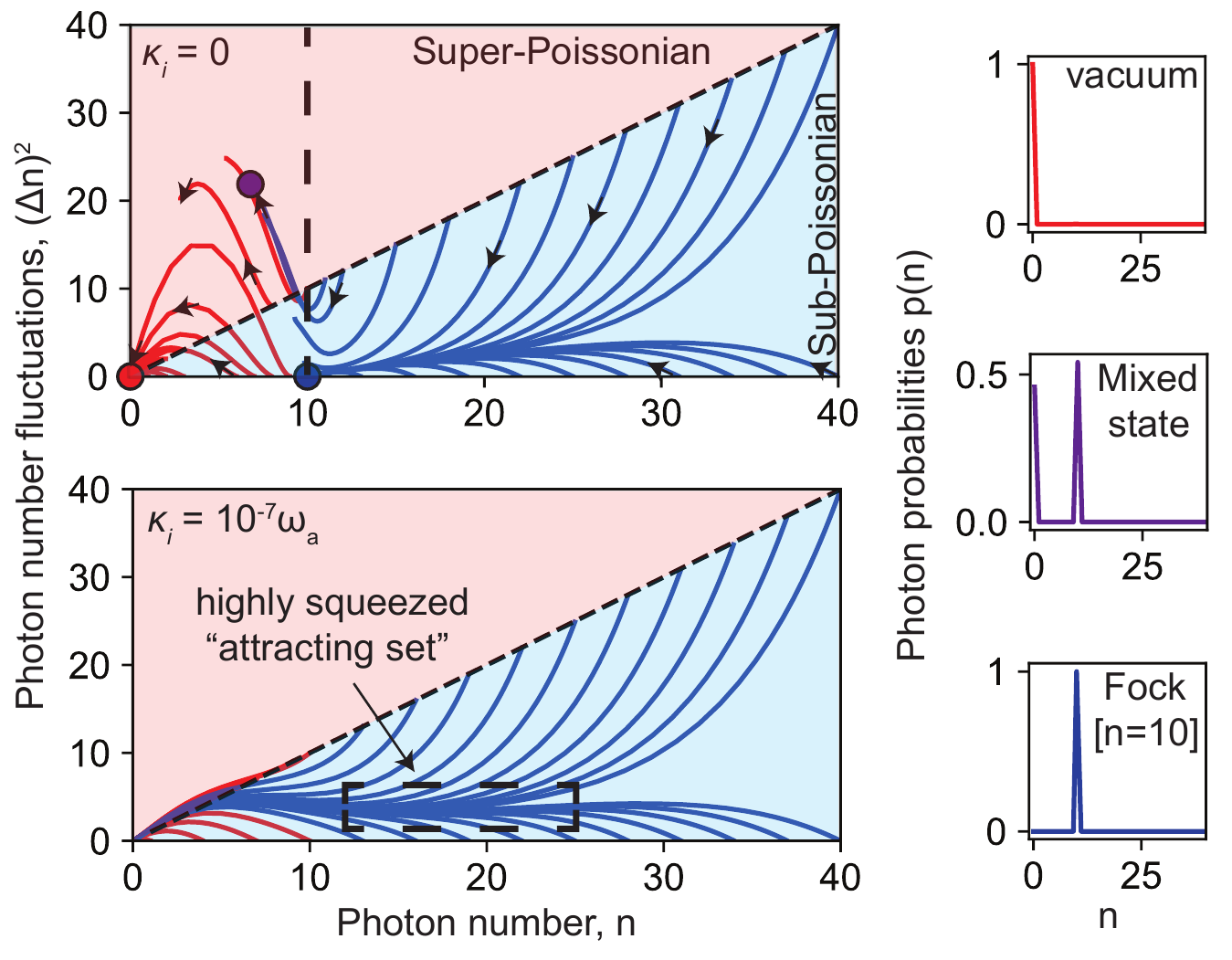}
    \caption{\textbf{Fock state generation and squeezing in a cavity undergoing nonlinear radiation loss.} Trajectories of the mean and variance of the photon number of the resonator in Fig. 2, for different initial values of the mean and variance. Poissonian states (points on the black diagonal) flow towards heavily number-squeezed (sub-Poissonian) states with variance much smaller than the mean. For long times, and for a system with an exact bound state in the continuum, rather than decaying to vacuum, they tend to Fock states (here, with $n_0 = 10$). For starting states with less than $10$ photons, they rapidly decay to vacuum states. Initial Fock states (points on bottom $x$-axis) start to have added noise (as usual), but then eventually experience a noise decrease and tend to $n=10$ Fock states. In the case of an external linear loss (here the external $Q$ is $5 \times 10^6$),the system does not tend to an exact Fock state, but impressive approximations are possible. For example, there are states (in the dashed black box) with noise 80\% below the classical limit, and with photon number uncertainty $\delta n = 2$ for mean photon number of 20. Parameters are the same as in Fig. 2. The state indicated by the purple circle is a mixed state in which there is some probability of the system being in a Fock state and some of it being in vacuum, which occurs when the initial probability distribution of the cavity photon number is nonnegligible both above and below $n_0 = 10$. }
    \label{fig:fock2}
\end{figure}
It is worth pausing to emphasize the striking-ness of this effect. In all known photonic systems, in the absence of a driving field, the only stable state is the vacuum state with zero photons. All finite photon-number states dissipate. In the systems examined here, \emph{all but two states dissipate}: the zero photon state, and the $n_0$-photon Fock state (the ``mixed'' state in Fig. 2 is a manifestation of that bistability). This type of bistability differs from conventional bistability in Kerr systems in that: with no driving amplitude, there is only a single stable state (vacuum) \cite{haus1984waves}. Further, in conventional bistability, Fock states do not arise \cite{drummond1980quantum}. 

In the presence of background losses, the BIC becomes imperfect and the bifurcation of trajectories in the top panel of Fig. 2 softens, as the state at $n = 10$ is no longer stuck -- it slowly leaks and progress towards the vacuum state. This is shown in the bottom panel for $\kappa_i = 10^{-7}\omega_a$ (an external $Q_i = 5 \times 10^6$). As can be seen, the blue trajectories start by initially moving well-below the black line, indicating that they become very strongly sub-Poissonian (number-squeezed). For example, if the initial state is Poissonian with a mean of 40 photons, at some point in time (about 300 ps in this example), the state turns into one with roughly 20 photons and an uncertainty $\Delta n = 2$, corresponding to a resonator state with fluctuations 80\% (about 7 dB) below the shot noise level, which is a very high degree of squeezing, especially for a state with a mesoscopic photon number like this one. 

\section{Prospects for experimental realization}

\begin{figure*}[ht!]
    \centering
    \includegraphics[width=1\textwidth]{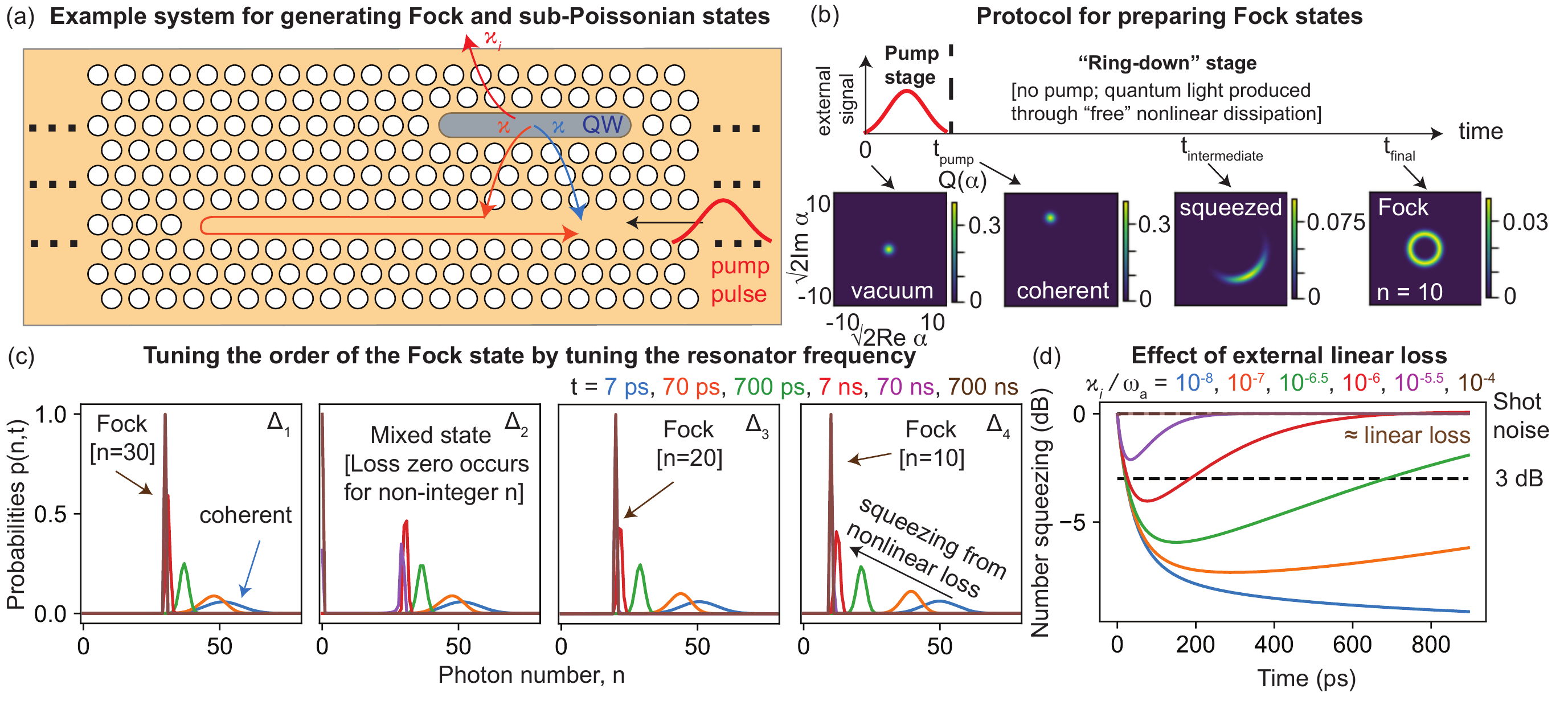}
    \caption{\textbf{Determinstic creation of large Fock states based on a pump-and-ringdown protocol in an example system: conversion of coherent states into Fock states.} (a) A system to realize bound states in the continuum and sizeable Kerr nonlinearity simultaneously: a photonic crystal slab with a (terminated) defect waveguide and a defect resonator, coupled to a semiconductor quantum well (light blue; on top of photonic crystal defect waveguide). (b) Pump-and-ringdown protocol for deterministic creation of Fock states. By sending in a short pulse, an initial coherent state of polaritons can be loaded. After the pump pulse, the system is left to evolve without any external pumping. This natural ``ring-down'', following Eq. (2), leads to the coherent state turning into a Fock state, as illustrated through the Husimi Q functions (see below for definition) for different times. (c) Evolution of the photon probabilities in time for the same initial state (a coherent state of 50 photons). The Fock state order can be tuned by changing the resonator frequency, as can be seen from the different panels, where we show tuning the Fock state order between 10 and 30 with near unit fidelity starting from a coherent state. Note that parameters for which $n_0$ in Fig. 2 is \emph{not} an integer (as in the second panel) lead to a ``failed'' Fock state (with substantial probability to be in vacuum - brown curve). Here, $\Delta_{1,2,3,4} \approx (0.0305,0.0300,0.0200,0.0095)\gamma$. (d) Photon-number squeezing as a function of time from injection of the pump pulse, for different levels of external linear loss. Importantly, even for realistic linear losses, it is possible to achieve very large number-squeezing, e.g., about 10 dB. For very large linear background loss (brown line), the result is indistinguishable from shot noise (0 dB squeezing). Parameters are the same as in Figs. 2 and 3. This figure further assumes an exciton-photon Rabi frequency of 1.8 meV, a nonlinear exciton-exciton energy of 20 $\mu \text{eV} \cdot \mu \text{m}^2$, and an exciton-photon hybridization coefficient of 0.5.  The input pulse preparing the coherent state is assumed to be 10 fs in duration and contain about 1000 photons (as from an attenuated pulsed laser). }
    \label{fig:fock3}
\end{figure*}

We have thus far addressed the physical principles behind the nonlinear loss, and why this generates Fock and squeezed states. We will now discuss concrete physical systems to implement the physics (and expected numbers), a protocol to deterministically prepare these states, and experimental signatures of them. The theory developed above is quite general, being applicable to any Kerr nonlinear oscillator coupled to one or more continuua with frequency-dependent couplings. As Kerr nonlinearities appear in many physical systems (photonics and beyond), the approach we take is to choose \emph{one} example system, and show the expected numbers from end-to-end (from pumping to detection), and provide a schematic discussion of other systems near the end.

The example we choose is meant to show how to generate close approximations of Fock states with large photon numbers ($n_0$ on the order of 10 or so, with $\Delta n \leq 1$). Such platforms, as per Eq. (6), should be realized using sizeable single-photon nonlinear strengths $\beta$. Fig. 4a illustrates such a system, formed by coupling excitons to a resonator which is then coupled to a waveguide \cite{wild2018quantum}. The parameters for the resonator-waveguide system are very similar to those in \cite{yu2021ultra}. The coupling between exciton and resonator leads to exciton-polaritons. Such excitations are the subject of many recent experiments (not limited to, but including \cite{boulier2014polariton,fink2018signatures,delteil2019towards,munoz2019emergence,ardizzone2022polariton}). From experiments, these exciton-polaritons are known to: (a)  reach strong coupling, with measured Rabi splittings exceeding the decay rate by 1-2 orders-of-magnitude (b) be well-described by a driven single-mode Kerr Hamiltonian (with damping), as in Eq. (1), with the lower polariton serving an ``effective photon'' (or photonic quasiparticle \cite{rivera2020light}) and (c) present the strong nonlinearities that are needed. The nonlinearities already present ($\beta \sim 10^{-5}\omega$) are already much larger than what is available in diffraction-limited microcavities of bulk nonlinear optical materials such as GaAs and GaP. In recent experiments, exciton-polaritons in microcavities have also been shown to present the characteristic optical bistability of Kerr systems, with concomitant squeezing \cite{boulier2014polariton}. Even more recently, it has been shown that polariton-polariton interactions are now strong enough to lead to antibunching of light, with promising prospects for photon blockade (or more appropriately, polariton blockade) upon improvement of the exciton lifetime (which in those experiments, was on the order of $10\text{ ps}$) \cite{fink2018signatures,delteil2019towards}. The most recent experiments have even managed to couple exciton polaritons in GaAs to optical bound states in the continuum in one-dimensionally periodic gratings, forming polariton BICs with measured lifetimes approaching 1 ns \cite{ardizzone2022polariton} (similar to a lower-bound associated with exciton dispersion, discussed for a different material platform \cite{wild2018quantum}). All in all, this suggests the use of such exciton-polaritons as a promising platform to realize the physics we describe here $-$ motivating its choice as our main example.

We now describe a protocol for ``loading'' Fock and highly squeezed states. It is illustrated in Fig. 4b: we start by injecting a short pulse through the resonator. This pulse is short compared to the timescale of the nonlinearity and the dissipation. Its purpose is to load an initial state with mean number of photons greater than $n_0$. The state can be somewhat arbitrary. Then, after the pulse passes, the dynamics are governed by the nonlinear dissipation of Eqs. (2-4). The simulated dynamics of the overall quantum state of the cavity, are visualized in Fig. 4b through the Husimi $Q$ functions ($Q(\alpha)=\langle \alpha|\rho|\alpha\rangle/\pi$), where $|\alpha\rangle$ is a coherent state. Initially, the state is vacuum. After the pump pulse, it is in a coherent state. The subsequent decay leads to a stretching of the $Q$ function in phase (angular direction), while the overall photon number decreases (this meniscus shape is well-documented in early works on Kerr squeezing \cite{andersen201630}). Then, after long times, the system approaches a Fock state. Fig. 4c shows the simulated photon probabilities, affirming the intuition of Fig. 1a. The main observation is that as a function of time, the photon noise of the coherent state \emph{condenses}: in other words, the nonlinear loss leads to photon number squeezing, until eventually, for longer times, the distribution converges to a Fock state (here, at times on the order of several hundred ns). But already, for shorter times (e.g., 700 ps or 7 ns), there is significant photon number squeezing (roughly 10 dB) and strong fidelity enhancement for Fock state generation. The limiting Fock state order depends on the detuning. For detunings which cause the nonlinear loss to vanish at an integer photon number, the Fock state is produced with fidelity 1. Meanwhile, for detunings for which the loss zero is non-integral, the resonator does not reach a Fock state, as there is no photon number for the distribution to get stuck at, leading to a ``failed'' Fock state.

Fig. 4d shows the effect of external (linear) losses on the intensity squeezing (it is the primary limitation to consider). The external loss is taken to come from the nonradiative decay of excitons. Fig. 4d shows the expected time-dependent photon-number-squeezing in the resonance (the lower-polariton) for different levels of external nonlinear loss. The main conclusions are that for external $Q$-factors in the range of $10^5-10^6$, there is substantial squeezing, even beyond the ``3 dB limit.'' A sufficiently large linear loss will fully mask the nonlinear loss (e.g., at $Q_i = 5 \times 10^3$), such that there is no noticeable deviation from what is expected from linear loss (shot noise stays as shot noise). The peak squeezing occurs around a few 100 ps. The current state of the art in exciton polaritons is on the order of 400 ps ($Q_i \approx 10^6$ $-$ near the green curve). This is about an order of magnitude away from a discussed bound associated with exciton dispersion (on the order of 1 ns exciton lifetime, corresponding to the orange curve), discussed for a different material platform \cite{wild2018quantum}. In other words, even within \emph{optics}, we already expect the new nonlinear dissipation we developed to be testable.

\textbf{Experimental signatures.} We now address the question of how to detect the generated quantum states. In Fig. 5, we show two possibilities: one standard technique, based on measuring second-order correlations, and one recently demonstrated technique which measures more directly the intracavity quantum state using a near-field probe (a free electron). 

\begin{figure}[t]
    \includegraphics[width=0.475\textwidth]{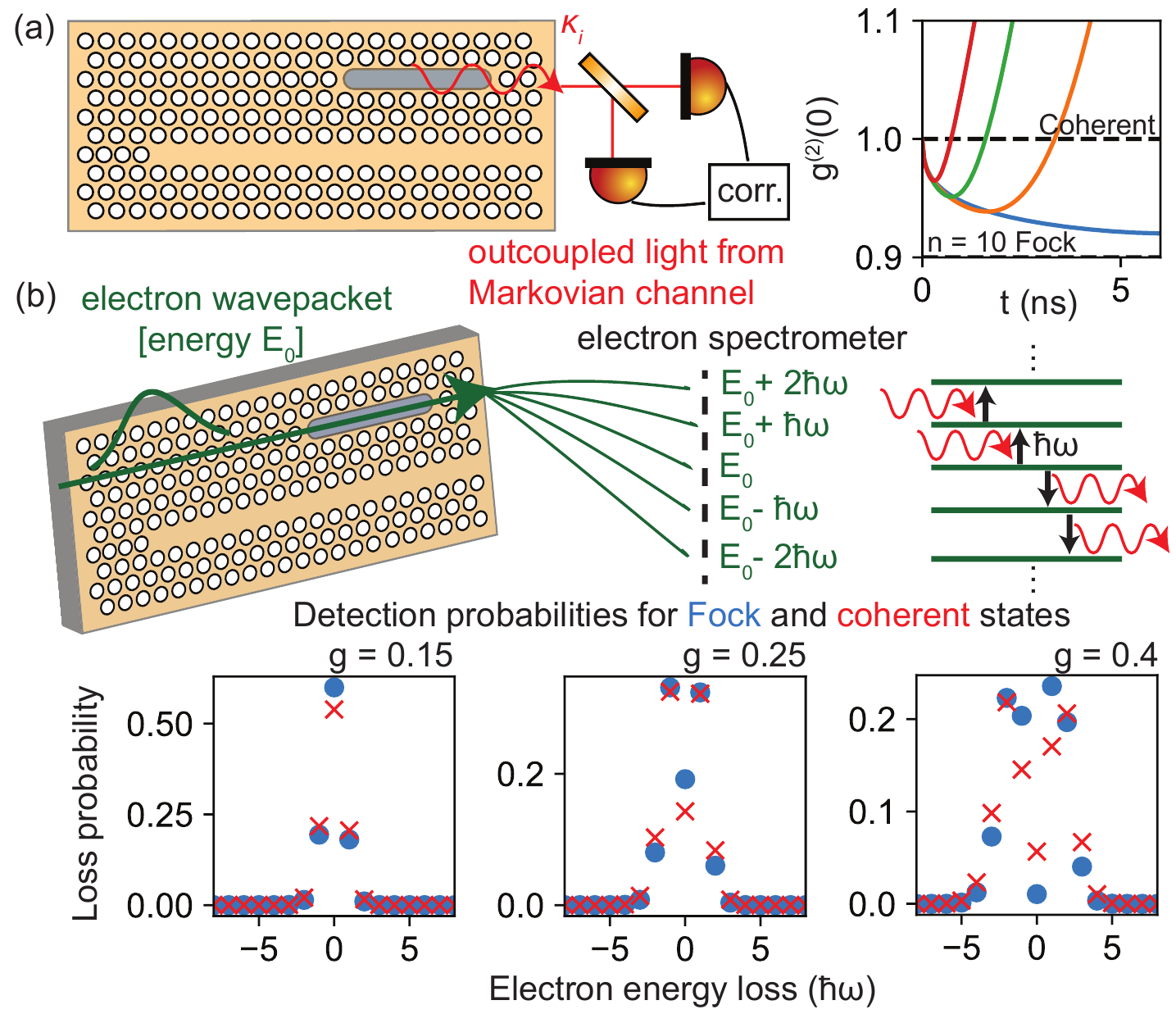}
    \caption{\textbf{Experimental signatures of intracavity Fock states.} (a) Far-field detection based on outcoupling light from a weak external outcoupler (e.g., vertically scattered light, or an additional outcoupling waveguide) and corresponding second-order correlations at zero time-delay, as measured by a Hanbury-Brown-Twiss interferometer. Colors in plot correspond to same external losses as in Fig. 4d. (b) Near-field detection based on quantized energy exchange between a free electron and the resonance via stimulated emission and absorption. Electron energy loss and gain probabilities are measured by an energy spectrometer for the electron, leading to energy losses which differ for Fock versus coherent states. Plots show energy loss probabilities for Fock and coherent states for three values of the electron-photon coupling strength (justification for values in SI) - indicating clear differences between Fock and coherent states. }
    \label{fig:fock2}
\end{figure}

Second-order correlation measurements have been extensively used to characterize quantum statistics of a resonance, even in systems of excitons coupled to microcavities (for example, one sends the light emitted from the resonator into a Hanbury-Brown-Twiss interferometer) \cite{fink2018signatures,munoz2019emergence, delteil2019towards}. Fig. 5a shows the expected second-order correlations (at zero time-delay) as a function of the measurement time, as well as the external losses. The Fock state of a given order will lead to $g^{(2)}(0) = \langle a^{\dagger 2}a^2 \rangle / \langle a^{\dagger}a \rangle^2 = 1-1/n$, so a Fock state of order $n=10$ will causes $g^{(2)}$ to approach 0.9. In the presence of external loss, the minimum $g^{(2)}$ increases towards 1. The above reveals a disadvantage of second-order correlation approaches. In particular, a $g^{(2)}$ between 0.9 and 1 could indicate a possible Fock state, or some other generic antibunched light state. Such techniques also appear unsuitable for larger Fock orders and intense squeezed light, where the deviation from unity would be much smaller than one. 

Recently, a technique has been demonstrated that is capable of measuring the quantum statistics of the intraresonator field with higher granularity. It is based on a new type of near-field photon detection technique, referred to as photon-induced near-field electron microscopy (PINEM). The basic theoretical description and experimental implementations are discussed in Refs. \cite{di2019probing,kfir2019entanglements,polman2019electron,rivera2020light,dahan2021imprinting}. The idea is illustrated in Fig. 5b: an energetic electron (typical kinetic energy $E_0 \sim$ 100-200 keV) grazes past the sample, interacting with the evanescent field of the resonance (it can also pass directly through the sample). The electron can then undergo absorption and stimulated emission induced by the photons in the resonance, leading to energy gain and loss. Due to the short duration of interaction ($<$ 1 ps), the electron probes the instantaneous density matrix of the light at delay time $\tau$. The electrons which pass through the sample are then sent through an electron spectrometer, which measures the energy loss or gain of the electron.

This technique in principle enables extraction of all normally-ordered moments of the field $(g^{(n)})$ through the electron loss probabilities \cite{dahan2021imprinting}. In this work, we are merely content to show how Fock states and coherent states (of the same average photon number) are differentiated. As one can see from Fig. 5b, there are clear differences (even for small $g$ such as 0.1) between the energy losses for Fock states and coherent states, which are noticeable for all energy losses with significant probability. That these differences are present over multiple data points (energy losses) is precisely what enables inversion to calculate moments beyond $g^{(2)}$, giving a characterization of the statistics directly of the cavity mode.

\section{Discussion and Outlook}

\begin{figure}[t]
    \centering
    \includegraphics[width=0.48\textwidth]{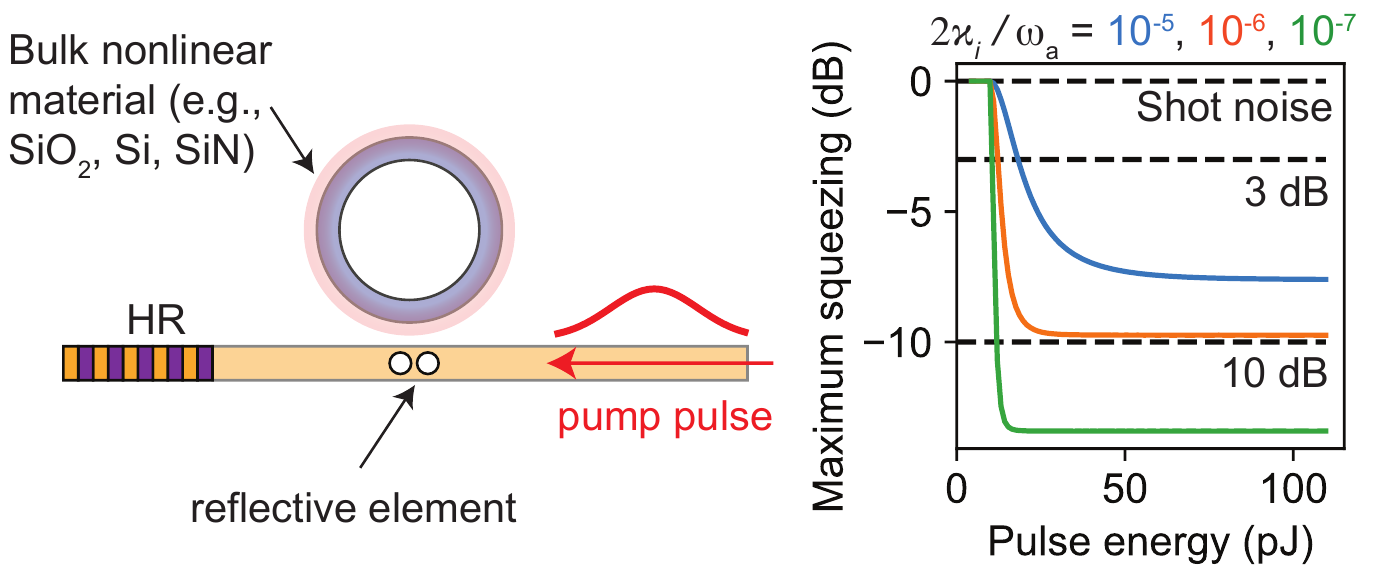}
    \caption{\textbf{Large photon-number squeezing using existing nanophotonic platforms.} (Left) a nanophotonic device consisting of a resonator (purple ring) and a partially reflective element (white circles), coupled to a waveguide which is terminated by a reflector (HR, e.g., a Bragg reflector). The ring and reflective element together realize a Fano resonance. When the resonator is loaded with photons by a pulse, the subsequent decay of those photons leads to development of extremely large squeezing in the resonator ($>$ 10 dB), which is plotted on the right as a function of the incident pulse energy and the limiting $Q = (2\kappa_i/\omega_a)^{-1}$ of the resonator (due to scatter and/or absorptive losses). Parameters taken here are $\omega_a = 0.6 \text{eV}$ (2 $\mu$m wavelength), $\beta = 10^{-10}\omega_a$, $2\kappa = 10^{-3}\omega_a$ (corresponding to a coupling $Q$ of $10^3$), $2\gamma = 10^{-2}\omega_a$ (corresponding to a distance between HR and resonator of 10 $\mu$m), $\Delta_0 = 10^{-1}\gamma$ and we assume an input pulse duration of 150 fs. }
    \label{fig:fock2}
\end{figure}

In optics, the realization of the nonlinear loss of Fig. 1 and the $n$-photon BIC effect predicted in Fig. 2 should already be within reach. Already Figs. 2 and 3 indicate possibilities for combining strong exciton-polariton nonlinearities with photonic bound states in the continuum to realize mesoscopic light states with $\Delta n \sim 1$. As discussed above, and also more extensively in SI Section IV, the nonlinearities are already strong enough, and low external losses have also very recently been achieved (exciton-polaritons have even already been interfaced with BICs as of this year).

Minimally, all that is required is a BIC (or a good approximation of a BIC) which appears for a certain index of refraction, and Kerr nonlinearity (third-order optical nonlinearity). The types of architectures to realize the former (illustrated in Fig. 1) have already been realized several times now (without Kerr nonlinearity). Fock states with extremely high fidelities would require sizable nonlinearities, but sufficiently interesting intensity squeezing should already be achievable in the macroscopic domain for bulk-type nonlinearities (e.g., in silicon or InGaAs). Nanophotonic systems more broadly (exploiting coupled cavities based on high-$Q$ ring resonators and microspheres \cite{vahala2003optical}, or photonic crystal cavities \cite{akahane2003high}) should enable the construction of almost arbitrary nonlinear losses, and even with very little background loss (though with weaker nonlinearities, leading to high squeezing rather than Fock state generation). Typically, on-chip nanophotonic squeezing has been limited by various external losses (e.g., associated with coupling out of the chip), but as of this year, significant progress has been made along these lines, achieving strong squeezing on-chip with second-order nonlinearities \cite{nehra2022few}. 

In Fig. 6, we back up this assertion by showing how existing Kerr nonlinear nanophotonic systems (where strong Kerr effects have been experimentally observed, see Refs. \cite{yang2020inverse,van2022all}), when illuminated by existing optical pulses (sub-nJ energies, 150fs duration), can lead to very large squeezings even for modest limiting quality factors. For example, if the limiting quality factor due to external loss is $10^5$, roughly 7 dB squeezing is possible, which would exceed any observation of nanophotonic squeezing if measured. More strikingly, for an attainable limiting $Q$, like $10^6$, over 10 dB is possible.

More extreme nonlinear dissipation, enabling one- and few-photon Fock states, could be achieved by combining these resonators with matter systems supporting single-photon-scale nonlinearities (e.g., cavity QED systems with photon blockade \cite{torma2014strong} or Rydberg atoms \cite{peyronel2012quantum} with BICs). 

Another worthwhile platform for implementing the physics described here is in superconducting circuits. Although several techniques already exist for creating Fock states in superconducting qubits (as discussed in the introduction, and reviewed in more depth in SI Section V), the approach we pose, which makes use only of Kerr nonlinearity and linear loss engineering, is quite flexible, and may be beneficial even when implemented in superconducting qubit systems $-$ it would enable for example the direct conversion of a microwave probe into a Fock state of a Kerr nonlinear microwave resonator (formed by coupling a Josephson junction with modest anharmonicity to a linear microwave resonator). From a nonlinearity and external loss perspective, we suspect the capabilities are more than present to demonstrate Fock-state and extreme squeezing with $n$-photon BICs. A simple heuristic argument is that even ten years ago, single-photon Kerr strengths in superconducting qubits were 30-times larger than the losses, which enables even cat-state generation from coherent states, one of the most exotic predictions in the quantum physics of the Kerr effect \cite{walls2007quantum}. Beyond providing a useful proving ground for the concepts developed here, our technique does provide a path to easily tune the Fock state order (just change the detuning), and achieve fairly high Fock-state numbers with high fidelity.

To summarize, we have presented a fundamentally new form of nonlinear dissipative interaction for photons. At the most basic level, the nonlinear dissipation arises from combining nonlinearity and leaky modes with frequency-dependent radiation loss. When the nonlinearity is Kerr, this combination induces a decay rate for photons with an intensity-dependence qualitatively beyond what is offered by commonly employed multiphoton and saturable absorbers. When the leaky-mode is an approximate BIC, the nonlinear dissipation creates a ``potential'' in photon number which facilitates the generation of Fock states and highly intensity-squeezed states.

As discussed earlier, the theory developed to describe such effects is quite general, as it is applicable to any Kerr nonlinear oscillator coupled to one or more continuua with frequency-dependent couplings. The consideration of Kerr nonlinearity is not so restrictive: many systems in nature with self-interactions are described by a Kerr Hamiltonian, with some value of the $\beta$ parameter which can be predicted from first-principles, or measured. Such systems include: bulk optical materials (where Kerr comes from $\chi^{(3)}$ \cite{boyd2020nonlinear}), exciton-polaritons (where Kerr comes from Coulomb interactions \cite{imamoglu1997strongly}), superconducting circuits (where Kerr comes from nonlinear inductance \cite{krantz2019quantum}), magnons (where Kerr comes from magnon-magnon interaction \cite{wang2018bistability}), Rydberg atoms (where single-photon nonlinearities arise from Rydberg blockade \cite{peyronel2012quantum}), and cavity-QED systems (where single-photon nonlinearities arise from photon blockade \cite{drummond1980quantum}). 

Our work establishes a new connection between two highly active fields: (1) radiation loss engineering, which has primarily been explored in classical optics in the context of BICs, exceptional points, and non-Hermitian photonics \cite{hsu2016bound,limonov2017fano,miri2019exceptional,leefmans2021topological,xia2021nonlinear} and (2) quantum-state engineering, where the use of nonlinear dissipation to engineer quantum states is well-appreciated (see e.g., Refs. \cite{puri2019stabilized,de2022error,harrington2022engineered}). In doing so, our work points to a new line of questions that we expect to be interesting in photonics and beyond. For example, beyond systems with BICs explored here, a natural subject to investigate would be quantum nonlinear systems with exceptional points, which are known to be sensitive to small changes in the refractive index \cite{chen2017exceptional}. Moreover, the general platform introduced here (nonlinearity plus frequency-dependent radiation loss) suggests the possibility of using second-order nonlinearity instead of Kerr. Since second-order nonlinearities enable phase-sensitive loss (and gain), it is clear that such systems enable qualitatively different opportunities. Such nonlinear losses, arising from second- and third-order nonlinearities might very well give paths towards stablizing other quantum optical states that are of interest to the community (Schrodinger cat states, GKP states, cluster states, and the like).

Given the generality and rich variety of effects introduced here, we expect the development of physical platforms to realize them may provide many exciting new areas for discovery in quantum optics, nonlinear optics, nanophotonics, and beyond.

\section{Acknowledgements} The authors acknowledge helpful and insightful discussions with Shiekh Zia Uddin, Charles Roques-Carmes, Hengyun Zhou, Ryotatsu Yanagimoto, Tatsuhiro Onodera, Logan Wright, Peter McMahon, Bertrand Halperin, Mikhail Lukin, Norman Yao, and Ido Kaminer. N.R. acknowledges the support of a Junior Fellowship from the Harvard Society of Fellows, as well as earlier support from a Computational Science Graduate Fellowship of the Department of Energy (DE-FG02-97ER25308), and a Dean's Fellowship from the MIT School of Science. J.S. was supported in part by the Department of Defense NDSEG fellowship no. F-1730184536. Y.S. acknowledges support from the Swiss National Science Foundation (SNSF) through the Early Postdoc Mobility Fellowship No. P2EZP2-188091. This material is based upon work supported in part by the Air Force Office of Scientific Research under the award number FA9550-20-1-0115; the work is also supported in part by the U. S. Army Research Office through the Institute for Soldier Nanotechnologies at MIT, under Collaborative Agreement Number W911NF-18-2-0048.

\bibliographystyle{unsrt}
\bibliography{fock.bib}

\begin{thebibliography}{10}

\bibitem{teich1989squeezed}
Malvin~C Teich and Bahaa~EA Saleh.
\newblock Squeezed state of light.
\newblock {\em Quantum Optics: Journal of the European Optical Society Part B},
  1(2):153, 1989.

\bibitem{davidovich1996sub}
Luiz Davidovich.
\newblock Sub-poissonian processes in quantum optics.
\newblock {\em Reviews of Modern Physics}, 68(1):127, 1996.

\bibitem{thomas2011real}
Nicholas Thomas-Peter, Brian~J Smith, Animesh Datta, Lijian Zhang, Uwe Dorner,
  and Ian~A Walmsley.
\newblock Real-world quantum sensors: evaluating resources for precision
  measurement.
\newblock {\em Physical review letters}, 107(11):113603, 2011.

\bibitem{braginskiui1975quantum}
Braginski{\u\i}.
\newblock Quantum-mechanical limitations in macroscopic experiments and modern
  experimental technique.

\bibitem{wang2020efficient}
Christopher~S Wang, Jacob~C Curtis, Brian~J Lester, Yaxing Zhang, Yvonne~Y Gao,
  Jessica Freeze, Victor~S Batista, Patrick~H Vaccaro, Isaac~L Chuang, Luigi
  Frunzio, et~al.
\newblock Efficient multiphoton sampling of molecular vibronic spectra on a
  superconducting bosonic processor.
\newblock {\em Physical Review X}, 10(2):021060, 2020.

\bibitem{aaronson2011computational}
Scott Aaronson and Alex Arkhipov.
\newblock The computational complexity of linear optics.
\newblock In {\em Proceedings of the forty-third annual ACM symposium on Theory
  of computing}, pages 333--342, 2011.

\bibitem{lund2014boson}
Austin~P Lund, Anthony Laing, Saleh Rahimi-Keshari, Terry Rudolph, Jeremy~L
  O’Brien, and Timothy~C Ralph.
\newblock Boson sampling from a gaussian state.
\newblock {\em Physical review letters}, 113(10):100502, 2014.

\bibitem{huh2015boson}
Joonsuk Huh, Gian~Giacomo Guerreschi, Borja Peropadre, Jarrod~R McClean, and
  Al{\'a}n Aspuru-Guzik.
\newblock Boson sampling for molecular vibronic spectra.
\newblock {\em Nature Photonics}, 9(9):615--620, 2015.

\bibitem{wang2017high}
Hui Wang, Yu~He, Yu-Huai Li, Zu-En Su, Bo~Li, He-Liang Huang, Xing Ding,
  Ming-Cheng Chen, Chang Liu, Jian Qin, et~al.
\newblock High-efficiency multiphoton boson sampling.
\newblock {\em Nature Photonics}, 11(6):361--365, 2017.

\bibitem{hamilton2017gaussian}
Craig~S Hamilton, Regina Kruse, Linda Sansoni, Sonja Barkhofen, Christine
  Silberhorn, and Igor Jex.
\newblock Gaussian boson sampling.
\newblock {\em Physical review letters}, 119(17):170501, 2017.

\bibitem{brod2019photonic}
Daniel~J Brod, Ernesto~F Galv{\~a}o, Andrea Crespi, Roberto Osellame,
  Nicol{\`o} Spagnolo, and Fabio Sciarrino.
\newblock Photonic implementation of boson sampling: a review.
\newblock {\em Advanced Photonics}, 1(3):034001, 2019.

\bibitem{zhong2020quantum}
Han-Sen Zhong, Hui Wang, Yu-Hao Deng, Ming-Cheng Chen, Li-Chao Peng, Yi-Han
  Luo, Jian Qin, Dian Wu, Xing Ding, Yi~Hu, et~al.
\newblock Quantum computational advantage using photons.
\newblock {\em Science}, 370(6523):1460--1463, 2020.

\bibitem{ourjoumtsev2007generation}
Alexei Ourjoumtsev, Hyunseok Jeong, Rosa Tualle-Brouri, and Philippe Grangier.
\newblock Generation of optical ‘schr{\"o}dinger cats’ from photon number
  states.
\newblock {\em Nature}, 448(7155):784--786, 2007.

\bibitem{heeres2017implementing}
Reinier~W Heeres, Philip Reinhold, Nissim Ofek, Luigi Frunzio, Liang Jiang,
  Michel~H Devoret, and Robert~J Schoelkopf.
\newblock Implementing a universal gate set on a logical qubit encoded in an
  oscillator.
\newblock {\em Nature communications}, 8(1):1--7, 2017.

\bibitem{hofheinz2008generation}
Max Hofheinz, EM~Weig, M~Ansmann, Radoslaw~C Bialczak, Erik Lucero, M~Neeley,
  AD~O’connell, H~Wang, John~M Martinis, and AN~Cleland.
\newblock Generation of fock states in a superconducting quantum circuit.
\newblock {\em Nature}, 454(7202):310--314, 2008.

\bibitem{rempe1990observation}
Gerhard Rempe, F~Schmidt-Kaler, and Herbert Walther.
\newblock Observation of sub-poissonian photon statistics in a micromaser.
\newblock {\em Physical review letters}, 64(23):2783, 1990.

\bibitem{varcoe2000preparing}
Benjamin~TH Varcoe, Simon Brattke, Matthias Weidinger, and Herbert Walther.
\newblock Preparing pure photon number states of the radiation field.
\newblock {\em Nature}, 403(6771):743--746, 2000.

\bibitem{sayrin2011real}
Cl{\'e}ment Sayrin, Igor Dotsenko, Xingxing Zhou, Bruno Peaudecerf, Th{\'e}o
  Rybarczyk, S{\'e}bastien Gleyzes, Pierre Rouchon, Mazyar Mirrahimi, Hadis
  Amini, Michel Brune, et~al.
\newblock Real-time quantum feedback prepares and stabilizes photon number
  states.
\newblock {\em Nature}, 477(7362):73--77, 2011.

\bibitem{uria2020deterministic}
Manuel Uria, P~Solano, and Carla Hermann-Avigliano.
\newblock Deterministic generation of large fock states.
\newblock {\em Physical Review Letters}, 125(9):093603, 2020.

\bibitem{canela2020bright}
VSC Canela and HJ~Carmichael.
\newblock Bright sub-poissonian light through intrinsic feedback and external
  control.
\newblock {\em Physical review letters}, 124(6):063604, 2020.

\bibitem{lounis2005single}
Brahim Lounis and Michel Orrit.
\newblock Single-photon sources.
\newblock {\em Reports on Progress in Physics}, 68(5):1129, 2005.

\bibitem{yamamoto1987generation}
Y~Yamamoto, S~Machida, N~Imoto, M~Kitagawa, and G~Bj{\"o}rk.
\newblock Generation of number-phase minimum-uncertainty states and number
  states.
\newblock {\em JOSA B}, 4(10):1645--1662, 1987.

\bibitem{waks2004direct}
Edo Waks, Eleni Diamanti, Barry~C Sanders, Stephen~D Bartlett, and Yoshihisa
  Yamamoto.
\newblock Direct observation of nonclassical photon statistics in parametric
  down-conversion.
\newblock {\em Physical review letters}, 92(11):113602, 2004.

\bibitem{cooper2013experimental}
Merlin Cooper, Laura~J Wright, Christoph S{\"o}ller, and Brian~J Smith.
\newblock Experimental generation of multi-photon fock states.
\newblock {\em Optics express}, 21(5):5309--5317, 2013.

\bibitem{ben2021shaping}
Adi Ben~Hayun, Ori Reinhardt, Jonathan Nemirovsky, Aviv Karnieli, Nicholas
  Rivera, and Ido Kaminer.
\newblock Shaping quantum photonic states using free electrons.
\newblock {\em Science Advances}, 7(11):eabe4270, 2021.

\bibitem{bonifacio1971quantum}
R~Bonifacio, P~Schwendimann, and Fritz Haake.
\newblock Quantum statistical theory of superradiance. i.
\newblock {\em Physical Review A}, 4(1):302, 1971.

\bibitem{bonifacio1971quantum2}
R~Bonifacio, P~Schwendimann, and Fritz Haake.
\newblock Quantum statistical theory of superradiance. ii.
\newblock {\em Physical Review A}, 4(3):854, 1971.

\bibitem{gonzalez2017efficient}
Alejandro Gonz{\'a}lez-Tudela, Vanessa Paulisch, HJ~Kimble, and J~Ignacio
  Cirac.
\newblock Efficient multiphoton generation in waveguide quantum
  electrodynamics.
\newblock {\em Physical Review Letters}, 118(21):213601, 2017.

\bibitem{kilin1995fock}
S~Ya Kilin and DB~Horoshko.
\newblock Fock state generation by the methods of nonlinear optics.
\newblock {\em Physical review letters}, 74(26):5206, 1995.

\bibitem{leonski1997fock}
W~Leo{\'n}ski, S~Dyrting, and R~Tana{\'s}.
\newblock Fock states generation in a kicked cavity with a nonlinear medium.
\newblock {\em journal of modern optics}, 44(11-12):2105--2123, 1997.

\bibitem{yanagimoto2019adiabatic}
Ryotatsu Yanagimoto, Edwin Ng, Tatsuhiro Onodera, and Hideo Mabuchi.
\newblock Adiabatic fock-state-generation scheme using kerr nonlinearity.
\newblock {\em Physical Review A}, 100(3):033822, 2019.

\bibitem{schmitt1998photon}
S~Schmitt, J~Ficker, M~Wolff, F~K{\"o}nig, A~Sizmann, and Gerd Leuchs.
\newblock Photon-number squeezed solitons from an asymmetric fiber-optic sagnac
  interferometer.
\newblock {\em Physical review letters}, 81(12):2446, 1998.

\bibitem{shirasaki1990squeezing}
Masataka Shirasaki and Hermann~A Haus.
\newblock Squeezing of pulses in a nonlinear interferometer.
\newblock {\em JOSA B}, 7(1):30--34, 1990.

\bibitem{bergman1991squeezing}
Keren Bergman and HA~Haus.
\newblock Squeezing in fibers with optical pulses.
\newblock {\em Optics letters}, 16(9):663--665, 1991.

\bibitem{kitching1994amplitude}
J~Kitching, R~Boyd, A~Yariv, and Y~Shevy.
\newblock Amplitude noise reduction in semiconductor lasers with weak,
  dispersive optical feedback.
\newblock {\em Optics letters}, 19(17):1331--1333, 1994.

\bibitem{friberg1996observation}
SR~Friberg, S~Machida, MJ~Werner, A~Levanon, and Takaaki Mukai.
\newblock Observation of optical soliton photon-number squeezing.
\newblock {\em Physical review letters}, 77(18):3775, 1996.

\bibitem{andersen201630}
Ulrik~L Andersen, Tobias Gehring, Christoph Marquardt, and Gerd Leuchs.
\newblock 30 years of squeezed light generation.
\newblock {\em Physica Scripta}, 91(5):053001, 2016.

\bibitem{nehra2022few}
Rajveer Nehra, Ryoto Sekine, Luis Ledezma, Qiushi Guo, Robert~M. Gray, Arkadev
  Roy, and Alireza Marandi.
\newblock Few-cycle vacuum squeezing in nanophotonics.
\newblock {\em Science}, 377(6612):1333--1337, 2022.

\bibitem{hsu2016bound}
Chia~Wei Hsu, Bo~Zhen, A~Douglas Stone, John~D Joannopoulos, and Marin
  Solja{\v{c}}i{\'c}.
\newblock Bound states in the continuum.
\newblock {\em Nature Reviews Materials}, 1(9):1--13, 2016.

\bibitem{rybin2017supercavity}
Mikhail Rybin and Yuri Kivshar.
\newblock Supercavity lasing.
\newblock {\em Nature}, 541(7636):164--165, 2017.

\bibitem{koshelev2018asymmetric}
Kirill Koshelev, Sergey Lepeshov, Mingkai Liu, Andrey Bogdanov, and Yuri
  Kivshar.
\newblock Asymmetric metasurfaces with high-q resonances governed by bound
  states in the continuum.
\newblock {\em Physical review letters}, 121(19):193903, 2018.

\bibitem{limonov2017fano}
Mikhail~F Limonov, Mikhail~V Rybin, Alexander~N Poddubny, and Yuri~S Kivshar.
\newblock Fano resonances in photonics.
\newblock {\em Nature Photonics}, 11(9):543--554, 2017.

\bibitem{yu2021ultra}
Yi~Yu, Aurimas Sakanas, Aref~Rasoulzadeh Zali, Elizaveta Semenova, Kresten
  Yvind, and Jesper M{\o}rk.
\newblock Ultra-coherent fano laser based on a bound state in the continuum.
\newblock {\em Nature Photonics}, 15(10):758--764, 2021.

\bibitem{drummond1980quantum}
PD~Drummond and DF~Walls.
\newblock Quantum theory of optical bistability. i. nonlinear polarisability
  model.
\newblock {\em Journal of Physics A: Mathematical and General}, 13(2):725,
  1980.

\bibitem{gardiner2004quantum}
Crispin Gardiner, Peter Zoller, and Peter Zoller.
\newblock {\em Quantum noise: a handbook of Markovian and non-Markovian quantum
  stochastic methods with applications to quantum optics}.
\newblock Springer Science \& Business Media, 2004.

\bibitem{walls2007quantum}
Daniel~F Walls and Gerard~J Milburn.
\newblock {\em Quantum optics}.
\newblock Springer Science \& Business Media, 2007.

\bibitem{haus1984waves}
Hermann Haus.
\newblock Waves and fields in optoelectronics.
\newblock {\em Prentice-Hall, Inc.}, 1984.

\bibitem{fink2018signatures}
Thomas Fink, Anne Schade, Sven H{\"o}fling, Christian Schneider, and Ata{\c{c}}
  Imamoglu.
\newblock Signatures of a dissipative phase transition in photon correlation
  measurements.
\newblock {\em Nature Physics}, 14(4):365--369, 2018.

\bibitem{boulier2014polariton}
T~Boulier, M~Bamba, A~Amo, Claire Adrados, A~Lemaitre, E~Galopin, I~Sagnes,
  J~Bloch, C~Ciuti, E~Giacobino, et~al.
\newblock Polariton-generated intensity squeezing in semiconductor
  micropillars.
\newblock {\em Nature communications}, 5(1):1--7, 2014.

\bibitem{delteil2019towards}
Aymeric Delteil, Thomas Fink, Anne Schade, Sven H{\"o}fling, Christian
  Schneider, and Ata{\c{c}} {\.I}mamo{\u{g}}lu.
\newblock Towards polariton blockade of confined exciton--polaritons.
\newblock {\em Nature materials}, 18(3):219--222, 2019.

\bibitem{munoz2019emergence}
Guillermo Mu{\~n}oz-Matutano, Andrew Wood, Mattias Johnsson, Xavier Vidal,
  Ben~Q Baragiola, Andreas Reinhard, Aristide Lema{\^\i}tre, Jacqueline Bloch,
  Alberto Amo, Gilles Nogues, et~al.
\newblock Emergence of quantum correlations from interacting fibre-cavity
  polaritons.
\newblock {\em Nature materials}, 18(3):213--218, 2019.

\bibitem{joannopoulos2011photonic}
John~D Joannopoulos, Steven~G Johnson, Joshua~N Winn, and Robert~D Meade.
\newblock {\em Photonic Crystals: Molding the Flow of Light}.
\newblock Princeton University Press, 2011.

\bibitem{scully1999quantum}
Marlan~O Scully and M~Suhail Zubairy.
\newblock Quantum optics, 1999.

\bibitem{fan2002analysis}
Shanhui Fan and John~D Joannopoulos.
\newblock Analysis of guided resonances in photonic crystal slabs.
\newblock {\em Physical Review B}, 65(23):235112, 2002.

\bibitem{fan2003temporal}
Shanhui Fan, Wonjoo Suh, and John~D Joannopoulos.
\newblock Temporal coupled-mode theory for the fano resonance in optical
  resonators.
\newblock {\em JOSA A}, 20(3):569--572, 2003.

\bibitem{gardiner1987input}
CW~Gardiner, AS~Parkins, and MJ~Collett.
\newblock Input and output in damped quantum systems. ii. methods in
  non-white-noise situations and application to inhibition of atomic phase
  decays.
\newblock {\em JOSA B}, 4(10):1683--1699, 1987.

\bibitem{wild2018quantum}
Dominik~S Wild, Ephraim Shahmoon, Susanne~F Yelin, and Mikhail~D Lukin.
\newblock Quantum nonlinear optics in atomically thin materials.
\newblock {\em Physical review letters}, 121(12):123606, 2018.

\bibitem{ardizzone2022polariton}
V~Ardizzone, F~Riminucci, S~Zanotti, A~Gianfrate, M~Efthymiou-Tsironi,
  DG~Su{\`a}rez-Forero, F~Todisco, M~De~Giorgi, D~Trypogeorgos, G~Gigli, et~al.
\newblock Polariton bose--einstein condensate from a bound state in the
  continuum.
\newblock {\em Nature}, 605(7910):447--452, 2022.

\bibitem{rivera2020light}
Nicholas Rivera and Ido Kaminer.
\newblock Light--matter interactions with photonic quasiparticles.
\newblock {\em Nature Reviews Physics}, 2(10):538--561, 2020.

\bibitem{di2019probing}
Valerio Di~Giulio, Mathieu Kociak, and F~Javier~Garc{\'\i}a de~Abajo.
\newblock Probing quantum optical excitations with fast electrons.
\newblock {\em Optica}, 6(12):1524--1534, 2019.

\bibitem{kfir2019entanglements}
Ofer Kfir.
\newblock Entanglements of electrons and cavity photons in the strong-coupling
  regime.
\newblock {\em Physical review letters}, 123(10):103602, 2019.

\bibitem{polman2019electron}
Albert Polman, Mathieu Kociak, and F~Javier Garc{\'\i}a~de Abajo.
\newblock Electron-beam spectroscopy for nanophotonics.
\newblock {\em Nature materials}, 18(11):1158--1171, 2019.

\bibitem{dahan2021imprinting}
Raphael Dahan, Alexey Gorlach, Urs Haeusler, Aviv Karnieli, Ori Eyal, Peyman
  Yousefi, Mordechai Segev, Ady Arie, Gadi Eisenstein, Peter Hommelhoff, et~al.
\newblock Imprinting the quantum statistics of photons on free electrons.
\newblock {\em Science}, 373(6561):eabj7128, 2021.

\bibitem{vahala2003optical}
Kerry~J Vahala.
\newblock Optical microcavities.
\newblock {\em nature}, 424(6950):839--846, 2003.

\bibitem{akahane2003high}
Yoshihiro Akahane, Takashi Asano, Bong-Shik Song, and Susumu Noda.
\newblock High-q photonic nanocavity in a two-dimensional photonic crystal.
\newblock {\em nature}, 425(6961):944--947, 2003.

\bibitem{yang2020inverse}
Ki~Youl Yang, Jinhie Skarda, Michele Cotrufo, Avik Dutt, Geun~Ho Ahn, Mahmoud
  Sawaby, Dries Vercruysse, Amin Arbabian, Shanhui Fan, Andrea Al{\`u}, et~al.
\newblock Inverse-designed non-reciprocal pulse router for chip-based lidar.
\newblock {\em Nature Photonics}, 14(6):369--374, 2020.

\bibitem{van2022all}
K~Van~Gasse, M~Cotrufo, KY~Yang, A~Al{\`u}, and J~Vuckovic.
\newblock All optical switching in a silicon nonlinear fano resonator.
\newblock In {\em CLEO: Science and Innovations}, pages SM4O--2. Optica
  Publishing Group, 2022.

\bibitem{torma2014strong}
P{\"a}ivi T{\"o}rm{\"a} and William~L Barnes.
\newblock Strong coupling between surface plasmon polaritons and emitters: a
  review.
\newblock {\em Reports on Progress in Physics}, 78(1):013901, 2014.

\bibitem{peyronel2012quantum}
Thibault Peyronel, Ofer Firstenberg, Qi-Yu Liang, Sebastian Hofferberth,
  Alexey~V Gorshkov, Thomas Pohl, Mikhail~D Lukin, and Vladan Vuleti{\'c}.
\newblock Quantum nonlinear optics with single photons enabled by strongly
  interacting atoms.
\newblock {\em Nature}, 488(7409):57--60, 2012.

\bibitem{boyd2020nonlinear}
Robert~W Boyd.
\newblock {\em Nonlinear optics}.
\newblock Academic press, 2020.

\bibitem{imamoglu1997strongly}
A~Imamoḡlu, Helmut Schmidt, Gareth Woods, and Moshe Deutsch.
\newblock Strongly interacting photons in a nonlinear cavity.
\newblock {\em Physical Review Letters}, 79(8):1467, 1997.

\bibitem{krantz2019quantum}
Philip Krantz, Morten Kjaergaard, Fei Yan, Terry~P Orlando, Simon Gustavsson,
  and William~D Oliver.
\newblock A quantum engineer's guide to superconducting qubits.
\newblock {\em Applied Physics Reviews}, 6(2):021318, 2019.

\bibitem{wang2018bistability}
Yi-Pu Wang, Guo-Qiang Zhang, Dengke Zhang, Tie-Fu Li, C-M Hu, and JQ~You.
\newblock Bistability of cavity magnon polaritons.
\newblock {\em Physical review letters}, 120(5):057202, 2018.

\bibitem{miri2019exceptional}
Mohammad-Ali Miri and Andrea Alu.
\newblock Exceptional points in optics and photonics.
\newblock {\em Science}, 363(6422), 2019.

\bibitem{leefmans2021topological}
Christian Leefmans, Avik Dutt, James Williams, Luqi Yuan, Midya Parto, Franco
  Nori, Shanhui Fan, and Alireza Marandi.
\newblock Topological dissipation in a time-multiplexed photonic resonator
  network.
\newblock {\em Nature Physics}, 2022.

\bibitem{xia2021nonlinear}
Shiqi Xia, Dimitrios Kaltsas, Daohong Song, Ioannis Komis, Jingjun Xu,
  Alexander Szameit, Hrvoje Buljan, Konstantinos~G Makris, and Zhigang Chen.
\newblock Nonlinear tuning of pt symmetry and non-hermitian topological states.
\newblock {\em Science}, 372(6537):72--76, 2021.

\bibitem{puri2019stabilized}
Shruti Puri, Alexander Grimm, Philippe Campagne-Ibarcq, Alec Eickbusch,
  Kyungjoo Noh, Gabrielle Roberts, Liang Jiang, Mazyar Mirrahimi, Michel~H
  Devoret, and Steven~M Girvin.
\newblock Stabilized cat in a driven nonlinear cavity: a fault-tolerant error
  syndrome detector.
\newblock {\em Physical Review X}, 9(4):041009, 2019.

\bibitem{de2022error}
Brennan de~Neeve, Thanh-Long Nguyen, Tanja Behrle, and Jonathan~P Home.
\newblock Error correction of a logical grid state qubit by dissipative
  pumping.
\newblock {\em Nature Physics}, 18(3):296--300, 2022.

\bibitem{harrington2022engineered}
Patrick~M Harrington, Erich~J Mueller, and Kater~W Murch.
\newblock Engineered dissipation for quantum information science.
\newblock {\em Nature Reviews Physics}, pages 1--12, 2022.

\bibitem{chen2017exceptional}
Weijian Chen, {\c{S}}ahin Kaya~{\"O}zdemir, Guangming Zhao, Jan Wiersig, and
  Lan Yang.
\newblock Exceptional points enhance sensing in an optical microcavity.
\newblock {\em Nature}, 548(7666):192--196, 2017.

\end{thebibliography}

\end{document}